\documentclass[sigconf]{acmart}  

\usepackage{enumerate}
\usepackage{multirow}
\usepackage{comment}
\usepackage{todonotes}

\usepackage{soul}
\usepackage{xcolor}
\newif\ifhighlight
\highlightfalse
\ifhighlight
\else
\renewcommand{\hl}[1]{#1}
\fi

\newcommand{\etal}{\textit{et al.}}
\newcommand{\eg}{\textit{e.g.}}
\newcommand{\ie}{\textit{i.e.}}

\AtBeginDocument{%
  \providecommand\BibTeX{{%
    \normalfont B\kern-0.5em{\scshape i\kern-0.25em b}\kern-0.8em\TeX}}}

\copyrightyear{2024}
\acmYear{2024}
\setcopyright{rightsretained}
\acmConference[CHI '24]{Proceedings of the CHI Conference on Human Factors in Computing Systems}{May 11--16, 2024}{Honolulu, HI, USA}
\acmBooktitle{Proceedings of the CHI Conference on Human Factors in Computing Systems (CHI '24), May 11--16, 2024, Honolulu, HI, USA}
\acmDOI{10.1145/3613904.3642870}
\acmISBN{979-8-4007-0330-0/24/05}

\begin{document}

\title{Swarm Body: Embodied Swarm Robots}

\author{Sosuke Ichihashi}
\affiliation{%
  \institution{Georgia Institute of Technology}
  \city{Atlanta}
  \state{Georgia}
  \country{USA}
  \postcode{30332}
}
\email{sichihashi3@gatech.edu}
\authornote{This work was completed on an internship at OMRON SINIC X Corporation.}

\author{So Kuroki}
\affiliation{%
  \institution{OMRON SINICX Corporation}
  \city{Tokyo}
  \country{Japan}
  \postcode{113-0033}
}
\email{so.kuroki@sinicx.com}

\author{Mai Nishimura}
\affiliation{%
  \institution{OMRON SINICX Corporation}
  \city{Tokyo}
  \country{Japan}
  \postcode{113-0033}
}
\email{mai.nishimura@sinicx.com}

\author{Kazumi Kasaura}
\affiliation{%
  \institution{OMRON SINICX Corporation}
  \city{Tokyo}
  \country{Japan}
  \postcode{113-0033}
}
\email{kazumi.kasaura@sinicx.com}

\author{Takefumi Hiraki}
\affiliation{%
  \institution{Cluster Metaverse Lab}
  \city{Tokyo}
  \country{Japan}
  \postcode{141-0031}
}
\email{t.hiraki@cluster.mu}

\author{Kazutoshi Tanaka}
\affiliation{%
  \institution{OMRON SINICX Corporation}
  \city{Tokyo}
  \country{Japan}
  \postcode{113-0033}
}
\email{kazutoshi.tanaka@sinicx.com}

\author{Shigeo Yoshida}
\affiliation{%
  \institution{OMRON SINICX Corporation}
  \city{Tokyo}
  \country{Japan}
  \postcode{113-0033}
}
\email{shigeo.yoshida@sinicx.com}

\renewcommand{\shortauthors}{Ichihashi, et al.}

\newcommand{\DIFF}[1]{\textcolor{black}{#1}}
\begin{abstract}

  The human brain's plasticity allows for the integration of artificial body parts into the human body. Leveraging this, embodied systems realize intuitive interactions with the environment. We introduce a novel concept: embodied swarm robots. Swarm robots constitute a collective of robots working in harmony to achieve a common objective, in our case, serving as functional body parts. Embodied swarm robots can dynamically alter their shape, density, and the correspondences between body parts and individual robots. We contribute an investigation of the influence on embodiment of swarm robot-specific factors derived from these characteristics, focusing on a hand. Our paper is the first to examine these factors through virtual reality (VR) and real-world robot studies to provide essential design considerations and applications of embodied swarm robots. Through quantitative and qualitative analysis, we identified a system configuration to achieve the embodiment of swarm robots.

\end{abstract}



\begin{teaserfigure}
  \includegraphics[width=\linewidth]{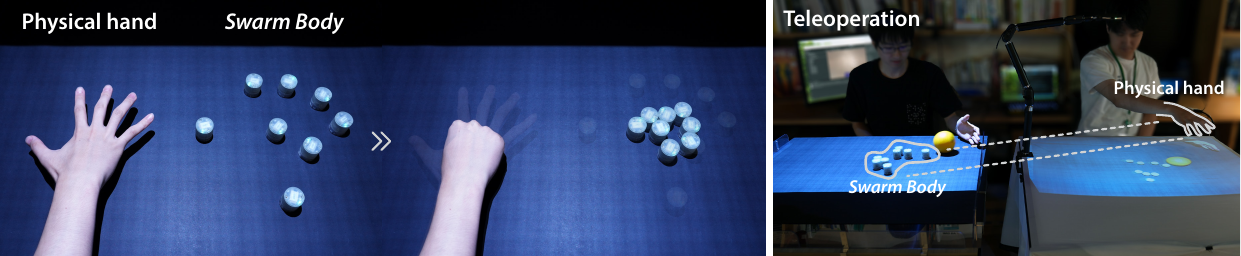}
  \caption{Embodied swarm robots allow a human to dynamically shape and disassemble their bodies to physically and adaptively interact with the local and remote environments. In the right figure, the user on the right pushes a ball toward the other person's hand with embodied swarm robots. The user can see the swarm robots projected on the table and control them in an embodied manner with the hand tracker attached to the table.}
  \Description{In the left figure, swarm robots form hand shapes according to the user's hand shapes. In the right figure, the robot operator watches a live projection of swarm robots on their desk and manipulates those robots by moving their own hands. The other person is physically interacting with those manipulated robots on a different desk by rolling a ball with them.}
  \label{fig:teaser}
\end{teaserfigure}


\maketitle

\section{Introduction}

The plasticity of the human brain allows it to recognize objects as body parts and manipulate them intuitively under certain conditions.
This cognitive ability has led to the development of embodied systems that enhance intuitive manipulation and immersion, such as teleoperated robotic arms or virtual avatars.
These systems have proven invaluable; however, there remains untapped potential in flexibility, adaptability to different environments, and overall robustness.
For example, it is difficult for conventional embodied systems to increase their levels of embodiment by dynamically changing their size to suit the individual, changing the shape to one suitable for a certain action, or for use on a tabletop or in an environment with obstacles.
To address these challenges, we introduce the concept of \emph{Embodied Swarm Robots}.

Swarm robots consist of many small robots of similar types, such as a school of fish or a swarm of insects.
Unlike conventional robotic arms, swarm robots are expected to cooperate and exert significant effects.
This unique quality endows them with robustness, flexibility, and scalability, enabling tasks such as navigation through narrow paths~\cite{roy2019Virtual}, pattern formation~\cite{alonso-mora2011Multirobot}, self-assembly~\cite{nakagaki2020HERMITS}, and the collective transport of objects that are larger and more complex than themselves~\cite{kuroki2023Collective}.
Thus, swarm robots have a wide range of applications such as environmental monitoring, space exploration, agriculture, emergency rescue, warehouses, industrial plants, entertainment, surveillance, and maintenance~\cite{schranz2020Swarm}.

Imagine a human body composed of \emph{swarms} of robots, offering unprecedented adaptability.
By using swarm characteristics, individuals can dynamically shape and equip their bodies to perform specific tasks and situations.
For example, a human with a swarm body can effortlessly traverse \hl{through} narrow paths accessible to the constituent robots of the swarm.
Moreover, if some individual robots within a swarm are lost, the remaining individuals compensate for the loss of functionality.
Embodying swarm robots holds promise for combining intuitive and immersive interactions with adaptability.
However, the realization of swarm robot body systems remains challenging because of our limited understanding of the conditions to support the sense of embodiment and subsequent system design considerations.

Although previous studies explored the embodiment of robot arms and virtual avatars, the embodiment of swarm robots introduces additional complexities.
\hl{For example, the size, density, and position distribution of robots, as well as the algorithms assigning the robots to the positions, can influence the levels of embodiment.}
Unlike robot arms, in which each moving link is constrained by joints, relative positions (\ie, position distributions) of swarm robots can change without geometrical constraints other than collisions, which gives them flexibility.
Thus, when representing a hand in a particular posture, various position distributions are possible, such as placing robots at the joint positions of the hand or placing them such that they are equally distributed within the hand shape.
In addition, the algorithm assigning robots to this position distribution influences the embodiment.
\hl{When moving a hand, the hand state (\ie, posture and position) changes dynamically.
Therefore, the robots need to follow each hand state while constantly assigning themselves to the position distribution for the current hand posture.
}
One possible assignment method is to \hl{statically} assign a particular robot to a particular \hl{position on a hand}, whereas another method is to dynamically update the assignments so that the total sum of the travel distances is minimized.
Therefore, identifying an appropriate algorithm for position distribution generation and assignment \hl{to follow the dynamically changing body states}, as well as the size and density of robots, is crucial for the successful embodiment of swarm robots.

Similar to the many previous studies on embodiment introduced in ~\autoref{sec:related_work}, this study investigated the embodiment of swarm robots by focusing on the hand, which is the part of the body where people most frequently interact with the environment.
In addition, we focused on tabletop swarm robots because we intended to explore various everyday interactions that people have with their hands at a table.
To evaluate the level of embodiment, we measured the sense of body ownership and agency~\cite{gallagher2000Philosophical}.
The factors examined were robot size, density, position distribution generation algorithm, and assignment algorithm.
VR and real-world robot experiments were conducted.
In the VR experiment, all the aforementioned factors were explored using simulated swarm robots.
This shows that swarm robots can be embodied and provides various insights into the embodiment of swarm robots with ideal swarm robot behavior.
However, actual robots may behave differently from those in VR environments.
To check whether swarm robots can be embodied and if similar embodiment characteristics are obtained in the real world, a similar embodiment experiment was conducted with the \hl{physical} swarm robots.
Based on the results of the VR and real-world experiments, we demonstrated the characteristics of swarm robot embodiment under ideal and actual robot behaviors, as well as design considerations for embodied swarm robot systems.
By comparing these results, we discuss how the characteristics of a real-world system affect the embodiment of swarm robots, thereby providing useful information for the future design of new embodied swarm robot systems.
Our contributions encompass:
\begin{enumerate}
    \item Proposing a framework for the embodiment of swarm robots in the hand.
    \item An algorithm to determine \hl{position distribution} relatively from the hand skeleton and dynamically assign them to robots, as determined through a series of VR and physical experiments, enhanced the sense of body ownership, \hl{sense of} agency, and overall usability compared with other conditions.
    \item Suggesting practical implementations and applications for embodied swarm robots that integrate these findings.
\end{enumerate}
\section{Related Work}
\label{sec:related_work}

\subsection{Body Ownership and Agency}
The sense of embodiment is a fundamental product of the human mind. 
A key aspect of embodiment is the sense of body ownership.
Body ownership is \textit{``the sense that I am the one who is undergoing an experience''}
~\cite{gallagher2000Philosophical}.
Botvinick and Cohen~\cite{botvinick1998Rubber} and others revealed that humans can perceive the sense of body ownership toward artificial objects with synchronous visuotactile stimulation.
In addition to synchronous visuotactile stimulation, synchronous visuomotor feedback also affects embodiment~\cite{sanchez-vives2010Virtual}.
Synchronous visuomotor feedback sheds light on another key aspect of embodiment: the sense of agency.
The sense of agency is \textit{``the sense that I am the one who is causing or generating an action''}~\cite{gallagher2000Philosophical}.
From the perspective of the sense of body ownership and agency, researchers have expanded their knowledge of various aspects of human embodiment, such as its temporal~\cite{sato2005Illusion, shimada2009Rubber, shibuya2018Relationship, dam2018Effects} and spatial~\cite{lloyd2007Spatial, kalckert2014spatial, romano2015robot, huynh2019Robotic} characteristics, as well as the embodiment possibilities of artificial bodies.

With advancements in VR technology, many scholars have explored the embodiment possibilities of various non-biological bodies, including elongated~\cite{kondo2018Invisible}, invisible~\cite{kondo2018Illusory}, discontinued~\cite{perez-marcos2012my, tieri2015Mere}, scrambled~\cite{kondo2019Scrambled}, re-associated~\cite{kondo2018Illusorya, kondo2020Reassociation}, shared~\cite{hagiwara2020Individuals, fribourg2021Virtual, takizawa2021Dynamic}, multiple~\cite{miura2021MultiSoma}, and supernumerary~\cite{ito2019We, arai2022Embodiment} body parts, as well as noncorporeal objects~\cite{ma2015Bodyownership}, through synchronous visuomotor feedback using VR.
These studies revealed that humans can embody not only their innate body parts with their original configurations but also those augmented or modified in certain ways as well as non-corporeal objects.
This knowledge can be applied in robotics to develop embodied robots that realize intuitive interactions, similar to human interactions with innate bodies.

\subsection{Robot Embodiment}
Robot embodiment has been studied for on-site and remote intuitive robot manipulations to interact with the environment physically.
For example, Aymerich-Franch \etal~ examined whether non-human--looking humanoid robot arms could be perceived as the user's arms and reported that the levels of embodiment were similar between human- and non-human--looking arms~\cite{aymerich-franch2017Nonhuman}.
Takada \etal~ showed that a single user could feel a sense of agency toward multiple robot arms while playing ping-pong with two opponents~\cite{takada2022Parallel}.
Most studies on virtual and robot embodiments have focused on body configuration similar to the human biological body.
Recently, researchers have begun to explore the embodiment of Supernumerary Robotic Appendages (SRAs) or unconventional body parts, such as additional limbs~\cite{sasaki2017MetaLimbs, umezawa2022Bodily, yamamura2023social}, hands~\cite{kieliba2021Robotic, yoshida2023tomura}, fingers~\cite{prattichizzo2014SixthFinger, shafti2021Playing}, joints~\cite{leigh2016Body}, and tails~\cite{ito2019We}.
The embodiment of an SRA or an unconventional body has the potential to augment human abilities and interactions with the environment, which are currently limited by the innate body configuration.
For example, tentacle limbs offer greater freedom than innate human limbs.
Similarly, swarm robots have the potential to provide robustness, flexibility, and scalability to a human body that conventional humans do not possess.
Therefore, this study is the first to investigate the embodiment of a swarm to realize embodied swarm robots.

\subsection{Swarm Robots in HCI}
Swarm robots operate such that many small robots of similar types cooperate with each other to exert significant effects.
This unique property endows the devices with robustness, flexibility, and scalability.
In HCI, actuated objects were used as tangible user interfaces (TUIs) to allow users to interact with the digital world through physical tabletop objects~\cite{pangaro2002actuated, frei2000curlybot, rosenfeld2004, sugimoto2011robotable2, richter2007, pendersen2007}.
Zooids~\cite{legoc2016Zooids} expanded these tangible interaction possibilities by introducing tabletop swarm robots and introduced the concept of Swarm User Interfaces (Swarm UIs).
This design space has been further extended by developing new robots~\cite{dementyev2016Rovables, lee2020Rolling}, adding new action possibilities to robots~\cite{zhao2017Robotic, suzuki2019ShapeBots, suzuki2021HapticBots, nakagaki2020HERMITS, lin2023ThrowIO, yu2023AeroRigUI}, building mixed reality system cooperative with projected images~\cite{hiraki2018, hiraki2019}, and introducing interfaces that seamlessly connect the digital and physical worlds~\cite{kaimoto2022Sketched, li2023Physica, ihara2023HoloBots, nakagaki2022Dis}.
Fundamental applications include sensing, visual feedback, haptic feedback, shape presentation, object representation, object actuation, environmental adaptation, and collaborative actions.
As such, the multi-agent nature of swarm robots has realized a flexible and scalable interface between the physical and digital worlds.

This multi-agent nature makes user control of swarm robots challenging.
Three main approaches have been adopted to control swarm robots: predefined, physical, and synchronized.
Many previous works have used predefined control in which a certain action or state triggers a certain robot's program.
Kim \etal~ compared various human-control modalities and strategies for predefined swarm robot control and provided guidelines~\cite{kim2020Userdefined}.
Another approach is to apply physical laws to robots so that users can expect and control their behavior~\cite{kaimoto2022Sketched, li2023Physica}.
Synchronized control, where swarm robots act in sync with corresponding remote or virtual objects, is also widely used, particularly in the context of TUIs, VR, and physical telepresence.
For example, HoloBots~\cite{ihara2023HoloBots} move such that their positions match the positions of the corresponding robots or the user's body part in the remote physical world.
These methods enabled physical interactions with the digital or remote physical worlds.
However, these swarm robots do not aim for embodiment or seek to represent human body parts.
By embodying swarm robots, the robot user can interact with the environment more intuitively and precisely while maintaining the power of the swarm robots.
In addition, the observer can understand the intentions of robot users more intuitively and precisely while feeling more humanness or affection toward the robots.
In human-robot interactions, an early attempt to control swarms in an embodied manner using fingers has already been made~\cite{kennel-maushart2023Interacting}; however, the robots do not represent the hand well, and their embodiment and interaction opportunities have not been investigated.
Inspired by these, this study explores the interaction opportunities of embodied swarm robots.
\section{Framework for Swarm Robots Embodiment}
\label{sec:framework}

\begin{figure*}[h]
    \centering
    \includegraphics[width=\linewidth]{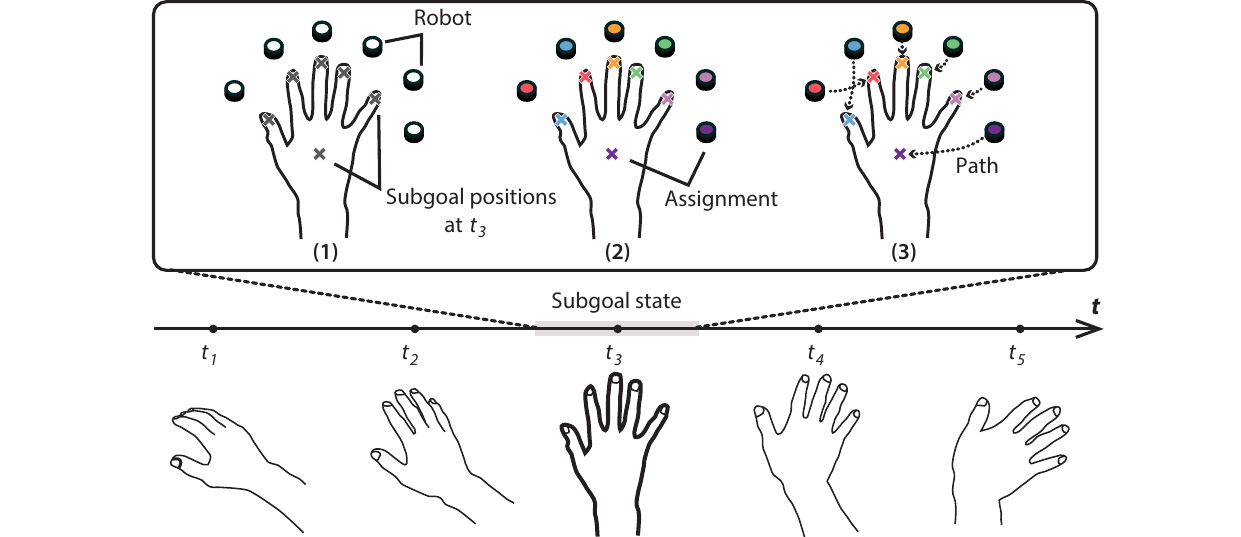}
    \caption{Example of the (1) subgoal formation generation, (2) robot assignment, and (3) local path planning.}
    \Description{The figure shows an example for each step of our framework for swarm robot embodiment. These are executed for each timestep. In the first step example, six subgoal positions are generated at the fingertips and the palm. In the second step example, these subgoal positions are matched to robots. In the their step example, a path for each robot to reach its goal is drawn.}
    \label{fig:steps}
\end{figure*}

We first introduce our framework to represent a hand with swarm robots in real-time.
\hl{As our body moves dynamically, navigating a swarm of robots has dynamic destinations associated with each timestep rather than a single static goal.
We refer to these dynamic destinations as subgoals that correspond to specific timesteps.
Assuming that the hand skeleton coordinates are tracked at regular intervals, we define the robot destination coordinates for each hand state as \emph{subgoal positions}.
If each robot moves to each subgoal position instantly, we can focus on how to represent each hand state with subgoal positions.
}

However, \hl{swarm robots often do not reach the subgoal positions before the subgoals are updated.
Therefore, it is necessary to control swarm robots to follow the subgoal positions so that the robots' movements represent the movement of the body part.}
Thus, to realize embodied motion with swarm robots, we require a new framework that simultaneously realizes both the proper hand representation and the smooth collective follow of swarm robots.

Common steps to control swarm robots are getting a \hl{subgoal} formation \hl{(a collection of subgoal positions)}, assigning the subgoal positions to the robots (assignment), and obtaining a path for each pair of robot\hl{s} and its subgoal position (path planning).
Inspired by this, we took the following three steps to achieve embodied movements of swarm robots as shown in \autoref{fig:steps}:
\begin{enumerate}
\item generate a \hl{subgoal} formation based on the current hand position and shape;
\item assign the generated subgoal positions to the robots;
\item obtain local paths and move the robots accordingly to avoid collisions with each other and obstacles.
\end{enumerate}

\subsection{Subgoal Formation Generation}
\label{subsec:subgoal_position_generation}

\begin{figure*}[h]
    \centering
    \includegraphics[width=\linewidth]{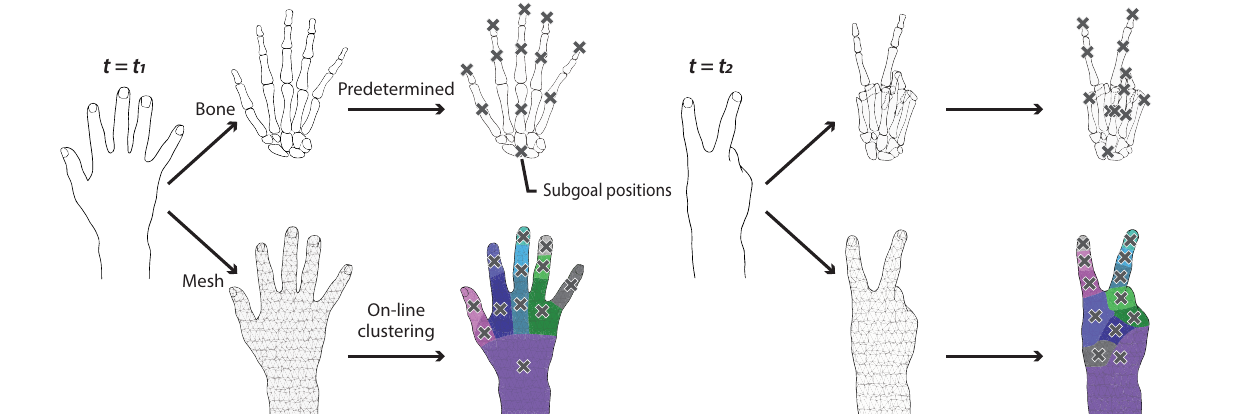}
    \caption{Bone- (top) and silhouette-based algorithms (bottom) to obtain subgoal formations. The bone-based algorithm obtains the subgoal formation based on the hand bone positions projected on the horizontal plane. The numbers in the figure correspond to the bone IDs provided by the Quest 2 hand tracking. The silhouette-based algorithm classifies the vertices of the hand mesh projected on the horizontal plane into clusters (shown in different colors in the figure) based on the number of robots. Then, the vertices closest to each cluster centroid become the subgoal positions.}
    \Description{The figure shows two examples for our algorithms to obtain subgoal formation for swarm robots. The left figure shows a paper hand shape example. The right figure shows a scissors hand shape example. In both of the figures, the hand bones in their hand shapes are shown, and twelve subgoal positions are plotted on the bones for the bone-based algorithm. For the silhouette-based algorithm, hand meshes are shown. The mesh vertices are categorized into twelve groups. The subgoal positions are plotted in each group.}
    \label{fig:distributions}
\end{figure*}

Step (1) sets the subgoal formation for the swarm robots.
Ideally, these positions are where the robots are; hence, this step affects the visual representation of the hand as well as the usability and interaction opportunity.

Because this step is similar to the virtual body representation, we took inspiration from that research.
We explored abstract virtual body representation approaches that are applicable to two-dimensional representation because we applied them to tabletop swarm robots.
Consequently, we identified two main approaches: point- and silhouette-based~\cite{wirth2021Impact}.
The point-based approach represents the body with a number of points corresponding to certain points on the body, whereas the silhouette-based approach represents a body with a solid shape in a single color.
In other words, the former focuses on specific points, and the latter focuses on overall shapes.
Based on these approaches, we designed two algorithms to obtain a set of subgoal positions from hand position, orientation, and shape, shown in \autoref{fig:distributions}.

\subsubsection{Bone-Based Subgoal Formation Generation}
Using the point-based approach, a set of points on the hand is determined, and the subgoal positions are moved according to the relative positions and movements of those points.
The set of points on the hand must be fixed with respect to the skeleton.
Thus, the bones of the hand were used as references for the tracked points.
Therefore, we refer to this as a bone-based algorithm.
Because the swarm robots we examined only moved on the horizontal plane, the bone positions were projected on the horizontal plane before calculating the subgoal positions.
For example, if the bone at the tip of the index finger is set as the subgoal, the robot moves to the bone position of the hand projected onto a horizontal plane.
\hl{The subgoal positions relative to the bones are predetermined (}\eg, \autoref{fig:vr_bone_locations} in our study).

\subsubsection{Silhouette-Based Subgoal Formation Generation}
Using the silhouette-based approach, the subgoal formation should reflect the overall shape of the hand.
To achieve this, a sensed hand skin mesh was obtained, and its vertices were clustered based on the number of required subgoal positions using the k-means algorithm~\cite{macqueenMETHODS}.
The subgoal positions were set to the vertex positions closest to the centroid of each cluster \hl{ensuring they remained within the hand outline.
The centroid itself may be placed outside the hand's outline, especially when clusters span multiple fingers, have gaps between fingers, or are around the proximal phalanges.}
We call this the silhouette-based algorithm.

We decided to examine both bone- and silhouette-based algorithms because they performed differently during our preliminary testing with two-dimensional hand representations using swarm robots.
The bone-based algorithm can offer more predictable robot movements to the user than the silhouette-based algorithm because the subgoal positions are always fixed to certain locations of the hand.
The silhouette-based algorithm constantly updates the subgoal positions on the hand.
Therefore, it does not guarantee that the subgoal positions are on the parts the user expects.
A potential advantage of the silhouette-based algorithm is its adaptability to various hand signs.
For example, when the user closes their hand, in the bone-based algorithm, the distance between the bones when projected onto the horizontal plane becomes so small that the subgoal positions become close, and the robots may collide.
However, in the silhouette-based algorithm, the subgoal positions are rarely too close to each other because clustering is conducted for that hand shape.

\subsection{Subgoal Position Assignment}
\label{subsec:subgoal_position_assignment}
Once the subgoal positions are obtained, they are assigned to the robots.
Two assignment methods are considered: \emph{static} and \emph{dynamic}.
The static method assigns a specific subgoal position to the same robot.
In dynamic assignment, the subgoal positions are constantly reassigned to realize smoother transitions from one hand shape to another.
This problem is defined as an assignment with variable subgoal formation.
This was attributed to the linear sum assignment problem~\cite{burkard2012Assignment}.
This problem can be solved by using the Hungarian algorithm~\cite{kuhn1955Hungarian}.

The \textit{bone-static} algorithm will offer more predictable robot movements than the other algorithm because the same robots always follow the same parts of the hand owing to the fixed subgoal formation and assignment.
By contrast, the dynamic assignment can avoid potential collisions of the robots.
For example, when the user turn\hl{s the hand from facing upward to facing downward}, dynamic assignment reassigns robots so that they do not have to flip their positions.

When the subgoal positions are constantly generated, and the current subgoal positions cannot be mapped to the past ones (\ie, when the silhouette-based subgoal formation is used), the static assignment cannot be applied because the same subgoal position does not exist at the next instant.
Therefore, the possible subgoal formation generation and assignment algorithms are: \textit{bone-static}, \textit{bone-dynamic}, and \textit{silhouette-dynamic}.

\subsection{Robot Control with A Local Path Planner}
Once pairs of a robot and subgoal position are determined, path planning and following robot control are executed. 
To plan paths and move multiple robots to their assigned subgoal positions, the Reciprocal Velocity Obstacles (RVO) algorithm~\cite{vandenberg2008Reciprocal} was used.
The RVO algorithm is an extension of the Velocity Obstacle concept, which offers navigation among passively moving objects by treating them as obstacles in the velocity space.
The RVO algorithm incorporates the assumption that other actively moving objects perform a similar collision avoidance behavior to the Velocity Obstacle and realized navigation among both passively and actively moving objects.
More precisely, the RVO algorithm with nonholonomic constraints~\cite{snape2010smooth} was used as most of the swarm robots, including the tabletop swarm robots we use, are nonholonomic.
\section{Embodiment Experiment in VR}
\label{sec:vr_study}
We first conducted a VR experiment to examine how various factors, including robot size, affect the levels of embodiment.
Then, to validate our findings in the real world, we conducted another embodiment experiment with fewer factors based on the VR results.
This experiment aimed to determine when swarm robots give users a sense of body ownership and agency.
Our experiment, designed based on prior embodiment studies, explored how specific swarm robot parameters affect body ownership, agency, and task load.
Participants interacted with virtual swarm robots, varying in size, density, and control algorithms.
After each trial, they answered questionnaires evaluating the sense of body ownership, sense of agency, and cognitive load.

\subsection{Participants}
The experiment involved 10 participants (6 males, 4 females; average age: 24.20 $\pm$ 2.57 SD).
Participants were sourced from a recruitment post on social media.
All participants were right-handed with normal or corrected vision and were unaware of the experiment's purpose.
Half of the participants had minimal VR experience, while the other half had extensive experience.
Participants signed a consent form regarding the experiment and were compensated with approximately \$16 in Amazon gift cards.
The ethics review board approved the experiment.

\subsection{Apparatus and Setup}

The experiment program, implemented using Unity, simulates and visualizes swarm robots and runs on a Windows-based computer.
A Meta Quest 2 HMD\footnote{\url{https://www.meta.com/quest/products/quest-2/}} tracked participants' hands and provided visual feedback based on the Unity visualization.
Participants wore noise-canceling headphones playing white noise to block external sounds.
An iPad collected the post-trial questionnaire responses.
In the VR, robots appeared on a table in front of the user as cylindrical bodies, a common form in HCI research~\cite{legoc2016Zooids}.
The subgoal formations are located on the table surface, and they are not translated in the horizontal direction from the participant's actual hand; \ie, there was no spatial discrepancy in the horizontal direction between the participant's hand and the robot's subgoal formation.

\subsection{Experiment Design}
\begin{figure}[h]
    \centering
    \includegraphics[width=\linewidth]{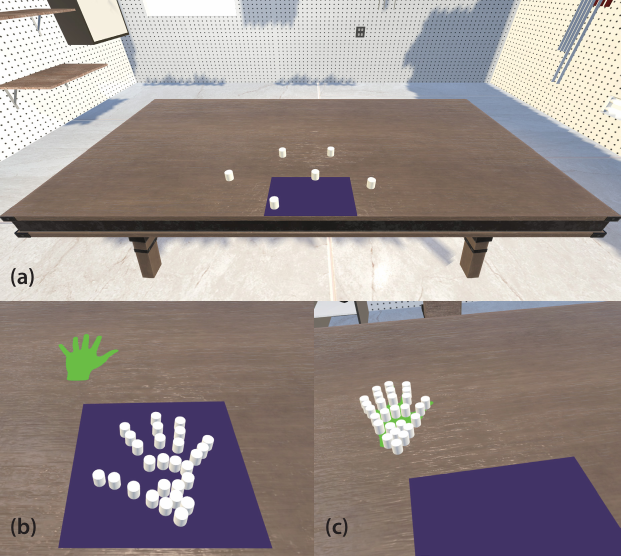}
    \caption{The VR setup for the reaching tasks (a). The purple area is the starting area, and the green object is the hand-shaped reaching target. Participants were instructed to (b) reach the target with the specified hand sign and (c) fit the swarm robots in the green area.}
    \Description{The top figure shows the overview of the virtual environment. It is a simple room with a wooden table at the center. The table has a purple zone near the participant's side. The bottom left figure shows a participant's viewpoint at the start of a trial. Swarm robots in the purple area form the paper hand shape. The target hand shape in green is shown near the top left corner of the table. The bottom right figure shows a participant's viewpoint at the end of a trial. The swarm robots are well-fitted on the green target in the paper hand shape.}
    \label{fig:vr_setup}
\end{figure}

\begin{figure}[h]
    \centering
    \includegraphics[width=\linewidth]{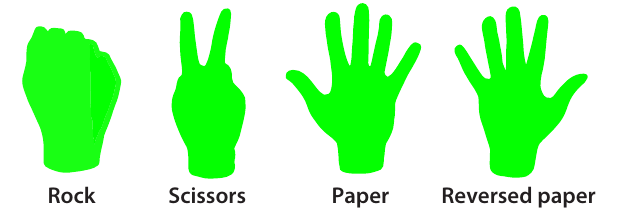}
    \caption{Hand signs used in the VR embodiment experiment. Participants make the rock, scissors, and paper shapes with their right hands, palms down. For the reversed paper sign, the palm should face up.}
    \Description{The figure shows the hand signs used in the experiement: Rock, Scissors, Paper, and Reversed Paper.}
    \label{fig:vr_hand_signs}
\end{figure}

\hl{We examined parameters (independent variables) specific to swarm robots.
The potential design parameters include robot's latency, speed, acceleration, size, color, shape, density, and control algorithm.
As it is not feasible to examine all the parameters, we focused on some of the parameters that are more unique for swarm robots: size, density, and control algorithm.
We excluded latency from the examined parameters because the effect of delay on embodiment is not unique to swarm robots and has been investigated in visual-motor synchronicity~\mbox{\cite{shibuya2018Relationship}}.
In a system with program-robot communication, we should consider the effect on embodiment owing to the sum of the delay caused by the robot's movement performance and the delay caused by this communication.
The first delay was excluded from this experiment by setting the robot's wheel speed to 400~mm/s with which we did not see much latency in the pilot study.
The second delay was excluded by removing the latency in the program-robot communication.}

We used a factorial design with three factors: 2 levels of the robot's size, 3 levels of the density, and 3 levels of the control algorithm. 
The independent variables examined were robot's size (\textit{30~mm} and \textit{20 ~mm}), density (\textit{sparse}, \textit{medium}, and \textit{dense}), and subgoal position generation and assignment algorithm (\textit{bone-static}, \textit{bone-dynamic}, and \textit{\textit{silhouette-dynamic}}).
All variables were within participants.

The experiment was conducted in the virtual environment shown in \autoref{fig:vr_setup}.
Participants were tasked with guiding swarm robots to a target sheet while making specific hand signs using their right hand.
The targets were green hand-shaped sheets with variations shown in~\autoref{fig:vr_hand_signs}.
Participants were instructed to change their hand signs to the target shapes when reaching the targets.
\hl{
A variety of hand signs were provided so that participants could explore the pros and cons of the subgoal position generation and assignment algorithms in various situations.
For example, the silhouette-based algorithm may have fewer collisions than the bone-based algorithm for the rock and scissors hand signs, but they may not make a big difference for the paper hand sign.
In the preliminary testing, the static and dynamic subgoal assignment algorithms resulted in very different robot behaviors (\ie, the static assignment sometimes caused the robots to get stuck, while the dynamic assignment did not) when the hand was flipped over.
Therefore, the reversed paper hand sign was also provided.
}

Participants were instructed to move the swarm robots representing their right hand to the purple starting area at the beginning of each task by moving their right hand.
\hl{Once all the robots have stayed} in the starting area for two seconds, a green hand-shaped reaching target appeared either at the left-front or at the right-front of the starting area.
Specifically, the target appeared 300~mm to the rear and 173~mm to the side from the center of the starting area.
The participants were instructed to reach for the object with the hand sign and to fit the robots in the target area as fast as possible.
The target disappeared five seconds after the task started.
The participants then return their hands and robots to the starting area.

Each trial consisted of eight tasks, \ie, 4 hand signs $\times$ 2 reaching positions.
The order of the hand signs and positions was randomized.
At the end of the eighth task, the participants were asked to take off the HMD with a text instruction.

\subsubsection{Robot's Size}
The robot's size influences the visual feedback of the swarm robots.
Two robot size conditions were prepared: 20 and 30~mm diameter.
The bigger size was set to 30~mm using the size of Zooids~\cite{legoc2016Zooids}, an open-source swarm robot often used in HCI research, as a reference.
The smaller size was set to 20~mm because it is close to the average adult human finger width~\cite{buchholz1991Ellipsoidal}.

\subsubsection{Robot's Density}
\begin{table}
  \caption{Number of Robots for Each Size and Density Condition.}
  \Description{The table shows the number of robots used in each size and density. For 20 mm robots, the sparse condition had six, the medium had eighteen, and the dense had twenty seven robots. For 30 mm robots, the sparse had six, the medium had eight, and the dense had twelve robots.}
  \label{tab:num_robots}
  \begin{tabular}{rccc}
    \toprule
    &Sparse&Medium&Dense\\
    \midrule
    20~mm & 6 & 18 & 27\\
    30~mm & 6 & 8 & 12\\
  \bottomrule
\end{tabular}
\end{table}

\begin{figure}[h]
    \centering
    \includegraphics[width=\linewidth]{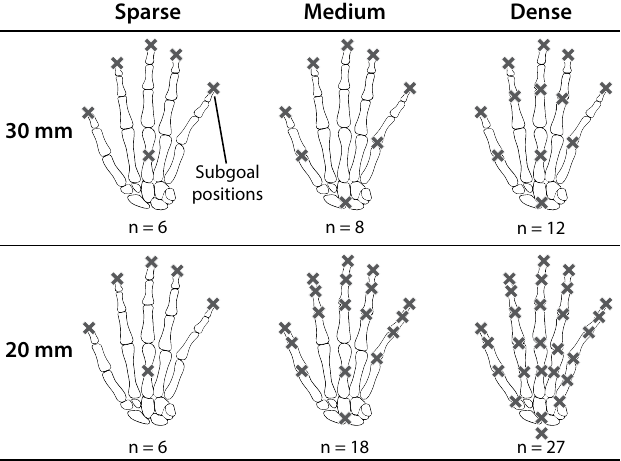}
    \caption{The predetermined subgoal positions relative to the hand bones. The gray crosses are the subgoal positions.}
    \Description{The figure shows the predetermined subgoal positions relative to the hand bones for each size and density condition. They were equally distributed over the hand while making sure there is a robot at each fingertip.}
    \label{fig:vr_bone_locations}
\end{figure}

The robot's density also affects the visual feedback of the swarm.
As the density increases, the spatial resolution of the visual representation increases.
However, this may complicate the representation and make control difficult as the number of robots increases.
Three density levels were considered: \textit{sparse}, \textit{medium}, and \textit{dense}.

We set the number of robots for each size and density condition based on preliminary testing.
The results are summarized in \autoref{tab:num_robots}.
Through the preliminary testing of different densities, we concluded that at least six robots representing each finger and palm were required to represent a hand, regardless of the robot's size.
Thus, we set the number of robots to six for the \textit{sparse} condition for both the 20~mm and 30~mm robots.
For the \textit{dense} condition, the number of robots was set to twelve for 30~mm robots because that is the maximum number of 30~mm robots that fit in an average adult hand.
To achieve the same density, the number of 20~mm robots was set to 27.
In the \textit{medium} condition, the number of 20~mm robots was set to 18, which was the average of the numbers in the \textit{sparse} and \textit{dense} conditions.
For 30~mm robots, the number for the \textit{medium} condition was set to eight to achieve the same density as the 20~mm robots in the \textit{medium} condition.

The subgoal positions relative to the hand bone need to be manually determined.
Thus, we designed the subgoal position distributions shown in \autoref{fig:vr_bone_locations} through the following step.
For the \textit{sparse} condition, we allocated subgoal positions to all the fingertips and the palm.
For the denser conditions, we increased the number of subgoal positions on the palm and distributed them around the palm.
If there are more subgoal positions, they were allocated to the proximal interphalangeal joints, to the metacarpophalangeal joints, and then, to the palm and the wrist.

\subsubsection{Subgoal Formation Generation and Assignment Algorithm}
As explained in \autoref{subsec:subgoal_position_generation} and \autoref{subsec:subgoal_position_assignment}, we examined three combinations of subgoal formation generation and assignment algorithms: \textit{bone-static}, \textit{bone-dynamic}, and \textit{silhouette-dynamic}.
The subgoal formation generation algorithm affects not only the visual feedback of the swarm but also the maneuverability of the subgoal formation.
Users using the bone-based algorithm can predict subgoal formation's movements more precisely than those using the silhouette-based algorithm.
This is because the silhouette-based algorithm continuously updates the correspondence between the hand and subgoal formation, and a certain hand movement does not necessarily result in the same subgoal formation's movements.

In addition, the assignment algorithm affects the robot's movements during hand shape transitions and hand movements.
It may further influence the user's understanding of the correspondence between the hand and robot movements.

\subsection{Measurements}
\begin{table*}
  \caption{Questionnaire used in the VR experiment.}
  \Description{The table shows the questionnaire items we used for measuring body ownership and agency.}
  \label{tab:vr_questionnaire}
  \begin{tabular}{ ll }
    \toprule
    Subscale&Questionnaire item\\
    \midrule
    \multirow{4}{8em}{Body ownership} & It felt like the swarm robot was my body. \\
                                    & It felt like some of the robots were my fingers. \\
                                    & It felt like the swarm robots belonged to me. \\
                                    & The swarm robot felt like a human hand. \\
    \midrule
    \multirow{4}{8em}{Agency} & The movements of the swarm robot felt like they were my movements. \\
                                    & I felt like I was controlling the movements of the swarm robot. \\
                                    & I felt like I was causing the movements of the swarm robot. \\
                                    & The movements of the swarm robot were in sync with my own movements. \\
  \bottomrule
\end{tabular}
\end{table*}

We used a questionnaire to assess participants' sense of body ownership and agency.
The questions related to the sense of body ownership and agency in the virtual embodiment questionnaire~\cite{roth2020Construction} were modified and used.
The questionnaire consisted of eight items in two subsets of questions for the sense of body ownership and agency, as shown in \autoref{tab:vr_questionnaire}.
Each response was scored on a seven-point Likert scale (\hl{1} = strongly disagree; \hl{7} = strongly agree).
The scores for the sense of body ownership and agency were calculated by taking the average of the corresponding four questions as suggested by Roth \etal~\cite{roth2020Construction}.
In addition, task load was measured using the NASA TLX questionnaire~\cite{hart1988Development}.
The pairwise comparisons of the factors were performed only after the first trial.

\subsection{Procedure}
Practice trials were conducted before the experiment to reduce learning effects.
Alcohol disinfection was performed on the experimental apparatus and the hands of the experimenter and participant.
Participants took a seat and were briefed on the experiment, procedures, data handling, risks, and rights and were instructed to sign a consent form if they agreed.
The participants put on the HMD, and if they had no problems with the fit, they did two practice trials (one with the \textit{30~mm}, \textit{sparse}, and \textit{bone-dynamic} settings and another with the \textit{30~mm}, \textit{dense}, and \textit{bone-dynamic}).
After the practice trials, the participants reviewed the questionnaires, and if they did not understand the question, an explanation was provided by the experimenter.

After the practice trials, the main experiment started.
Participants wore the HMD \hl{while making the four hand signs in ~\mbox{\autoref{fig:vr_hand_signs}} a random order, making each hand sign twice, once toward the left side of the table and once toward the right side of the table.}
\hl{Then, they} removed the HMD and answered the questionnaires on an iPad.
The participants repeated the task of making the hand signs and questionnaire responses a total of 18 times.
To control for learning effects, the order of the experimental conditions \hl{(robot size, density, and subgoal position generation and assignment algorithm)} was randomized.
To control the interference effect of arm fatigue, the participants were asked to ensure that they were not fatigued before each task.
A five-minute break was provided after the ninth task.
After the 18th questionnaire response, the participants filled out a demographic questionnaire, and a semi-structured interview was conducted for approximately five to ten minutes.
The entire experiment lasted approximately 90 min.

\subsection{Results}
\begin{figure}[h]
    \centering
    \includegraphics[width=\linewidth]{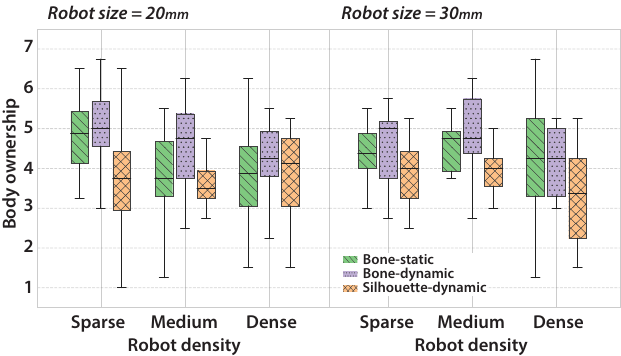}
    \caption{Body ownership score obtained in the VR experiment for each size, density, and subgoal formation generation and assignment algorithm.}
    \Description{The figure shows the body ownership score obtained in the VR experiment for each size, density, and subgoal formation generation and assignment algorithm.}
    \label{fig:vr_bo}
\end{figure}

\begin{figure}[h]
    \centering
    \includegraphics[width=\linewidth]{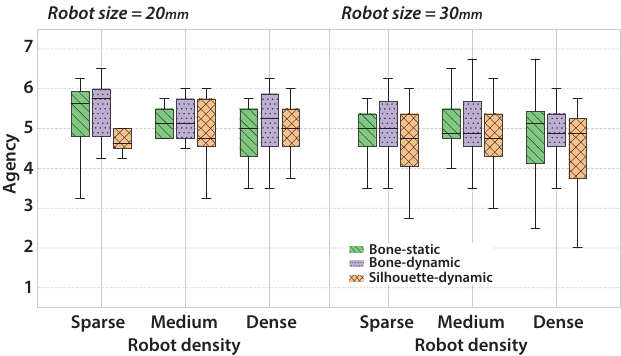}
    \caption{Agency score obtained in the VR experiment for each size, density, and subgoal formation generation and assignment algorithm.}
    \Description{The figure shows the agency score obtained in the VR experiment for each size, density, and subgoal formation generation and assignment algorithm.}
    \label{fig:vr_agency}
\end{figure}

\begin{figure}[h]
    \centering
    \includegraphics[width=\linewidth]{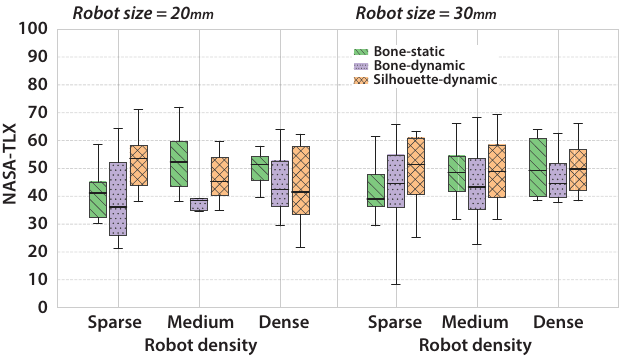}
    \caption{Task load index obtained in the VR experiment for each size, density, and subgoal formation generation and assignment algorithm.}
    \Description{The figure shows the task load index obtained in the VR experiment for each size, density, and subgoal formation generation and assignment algorithm.}
    \label{fig:vr_tlx}
\end{figure}

The body ownership score, agency score, and task load index were calculated for three factors: robot size, density, and subgoal formation generation and assignment algorithm.
The results are shown in \autoref{fig:vr_bo}, \autoref{fig:vr_agency}, and \autoref{fig:vr_tlx}.
As these were nonparametric data, we performed an Aligned Rank Transform (ART)~\cite{wobbrock2011aligned} followed by a three-factor two-way repeated-measure ANOVA with Holm correction for each subscale (\ie, body ownership, agency, and task load) to investigate the main effects and interactions.

\subsubsection{Body Ownership}
ANOVA revealed significant main effects of density ($F(2, 153)=4.00, $\hl{ $p=.020$}$, \eta_{p}^{2}=.05$) and subgoal formation generation and assignment algorithm on the body ownership score ($F(2, 153)=15.28, $\hl{ $p=.000$}$, \eta_{p}^{2}=.17$).
There was a trend toward an interaction between size and density, but no significant interaction was found between any of the factors.
Therefore, using the Holm-corrected ART-C~\cite{elkin2021Aligned}, contrast tests were performed on the size, density, and subgoal formation generation, and assignment algorithm factors.
The contrast test on the size factor revealed that there was no significant difference in body ownership score between \textit{20~mm} and \textit{30~mm} conditions.
The \textit{sparse} condition led to a significantly higher body ownership score than the \textit{dense} condition~($p=.016$, \hl{cohen's $d=0.516$}).
Finally, the \textit{bone-dynamic} resulted in a significantly higher body ownership score than the other two~(\textit{bone-dynamic} and \textit{silhouette-dynamic}: \hl{$p=.000$, cohen's $d=1.002$}, \textit{bone-dynamic} and \textit{bone-static}: \hl{$p=.033$, cohen's $d=0.393$}), and \textit{bone-static} resulted in a significantly higher body ownership score than the \textit{silhouette-dynamic}~(\hl{$p=.002$, cohen's $d=0.609$}).

\subsubsection{Agency}
ANOVA revealed significant main effects of size ($F(1, 153)=5.92, $\hl{ $p=.016$}$, \eta_{p}^{2}=.04$) and subgoal formation generation and assignment algorithm on the agency score ($F(2, 153)=5.07, $\hl{ $p=.007$}$, \eta_{p}^{2}=.06$).
No significant interactions were observed between any of the factors.
Therefore, using the Holm-corrected ART-C, contrast tests were performed on size, density, and subgoal formation generation and assignment algorithm factors.
The contrast test on the size factor showed that the \textit{20~mm} condition led to a significantly higher agency score than the \textit{30~mm} condition~(\hl{$p=.016$, cohen's $d=0.363$}).
In addition, the \textit{bone-dynamic} algorithm resulted in a significantly higher agency score than the \textit{silhouette-dynamic} algorithm~(\hl{$p=.006$, cohen's $d=0.576$}).
No significant differences in agency score were observed for the density factor.

\subsubsection{Cognitive Load}
The ANOVA revealed significant main effects of subgoal formation generation and assignment algorithm on cognitive load ($F(2, 153)=6.11, $\hl{ $p=.002$}$, \eta_{p}^{2}=.07$).
However, a significant interaction between density and the algorithm was found ($F(4, 153)=3.03, $\hl{ $p=.019$}$, \eta_{p}^{2}=.07$).
Therefore, multiple comparisons by ART-C (holm corrected) were performed.
As a result, no significant differences were found among all groups, but there was a trend toward differences between \textit{silhouette-dynamic} + \textit{sparse} and \textit{bone-static} + \textit{medium} as well as \textit{bone-static} + \textit{medium} and \textit{bone-static} + \textit{sparse}.
In addition, the \hl{difference in} differences test for the density--algorithm interaction showed significant differences between the \textit{silhouette-dynamic} - \textit{bone-static} and \textit{dense} -- \textit{sparse}~(\hl{$p=.039$, cohen's $d=1.278$}) as well as the \textit{silhouette-dynamic} -- \textit{bone-static} and \textit{medium} -- \textit{sparse}~(\hl{$p=.036$, cohen's $d=1.307$}).

\subsubsection{Semi-Structured Interview}
During the interview at the end of the experiment, a few common items were reported.
Nine participants reported that the swarm robot felt like their hand \hl{at least once}, and another participant reported that it felt like it was following their hand.
This one participant noted that they often felt that the robot moved late relative to their hand, which led to the sensation that robots were following their hand.
The other eight participants also noted that the swarm robots seemed slower under certain conditions and that increasing their speed would lead to a stronger embodiment.

\hl{Seven} participants noted the importance of fingertips in feeling \hl{like} the swarm robot as a hand.
They reported that when the robot was positioned for each fingertip, they tended to recognize and move it as a hand.
In addition, six stated that the hand-like appearance was lost when the robots collided or vibrated with each other, or when they moved, and when they did not fit well into the hand shape, coalesced around the palm, or were misaligned with the hand position.

All participants also mentioned the impact of the size of the robot.
Two stated that the larger robot felt more like a hand or that they felt in control, while five stated that the smaller robots felt more like a hand or fingers.

Nine participants mentioned the influence of robot density.
Eight stated that the lower density was more likely to cause the embodiment or lead to a higher maneuverability, but three of these stated that the higher density was felt as a hand when the robot's size was small, \ie, 20~mm.
Another participant \hl{felt like swarm robots were the hand} regardless of density.

\subsection{Discussion}
In a VR psychophysical experiment, we studied how swarm robot factors like size, density, and algorithm impact embodiment.
We focused on the sense of body ownership, sense of agency, and task load in the experimental results.

\subsubsection{Embodiment Across Different Sizes of Swarm Robots}
\label{subsubsec:_vr_discussion_size}
We first compared the embodiment scores for each size condition with the neutral level \hl{(\ie, 4 point rating, which is a neutral response to the 7-point Likert scale questionnaire) to evaluate the level of embodiment.
This neutral level was the null hypothesis of the tests.}
The body ownership scores were significantly higher than the neutral level for both \textit{30 mm} (\hl{$p=.018$, cohen's $d=0.413$}) and \textit{20 mm} (\hl{$p=.039$}, cohen's $d=0.351$), and the same was true for the agency scores (\textit{30 mm}: \hl{$p=.000$, cohen's $d=1.920$}; \textit{20 mm}: \hl{$p=.000$, cohen's $d=2.172$}).
\hl{
The cohen's $d$ values for body ownership score indicate small to medium effect sizes, and those for agency score indicate large effect sizes.
Thus, the questionnaire responses imply that the participants felt a higher level of embodiment for both \textit{30 mm} and \textit{20 mm} robots than the neutral state in which participants neither deny or affirm whether they feel the robots as their body parts or as if they are in control of the robots.
Although these alone do not robustly show that Swarm Body was embodied by the participants for both conditions, they are consistent with the interview responses that many participants felt swarm robots like their hands at least once.
Therefore, Swarm Body is likely to be embodied to some extent for both} \textit{30 mm} and \textit{20 mm} robots with an appropriate algorithm and density.

When comparing the size conditions, no significant differences were found between the \textit{20~mm} and \textit{30~mm} conditions in the body ownership score and task load index, while the agency score was significantly higher for \textit{20~mm} than for \textit{30~mm} with a small to medium effect size (cohen's $d=0.363$).
A possible explanation for this is that larger perceived movements of robots when moving fingers led to a higher sense of agency.
The movement of a robot relative to its size is larger for a smaller robot.
Thus, the movements of the 20~mm robots could be perceived as larger than those of the 30~mm robots.
The reported preferences for robot size also varied in the interviews.
Therefore, the impact of robot size on the embodiment and the resulting cognitive load is likely to vary among individuals, yet smaller robots tend to provide a higher sense of agency.

\subsubsection{Sparse Swarm Robots Achieve A Higher Sense of Body Ownership}
\label{subsubsec:vr_discussion_density}
The body ownership score was significantly higher in the \textit{sparse} than in the \textit{dense} with a medium effect size (cohen's $d=0.516$), suggesting that swarm robots with lower density is more likely to be felt as a hand.
This is consistent with participants' reports that the lower density of the swarm robots felt more like hands.
This may be due to the more frequent occurrence of collisions and vibratory movements of robots, their deviations from the hand shape, and getting stuck around the palm in the higher density conditions, which reduced the sense of embodiment as described by the participants.
Another possible reason is that the participant could tell which robots represent fingertips better in the \textit{sparse} condition; \ie, it offers a better understanding of finger-robot correspondence with a simpler representation.
This might result in a higher sense of body ownership as discussed in~\autoref{subsubsec:vr_discussion_algorithm}.

The body ownership score for the \textit{sparse} was also significantly higher than \hl{the neutral level ($p=.004$, cohen's $d=0.639$)}.
The agency scores were significantly higher than the neutral level in all the conditions~(\textit{dense}: \hl{$p=.000$, cohen's $d=1.676$}; \textit{medium}: \hl{$p=.000$, cohen's $d=1.918$}; \textit{sparse}: \hl{$p=.000$, cohen's $d=1.982$}), although there were no significant differences between the conditions.
\hl{As stated in~\mbox{\autoref{subsubsec:_vr_discussion_size}}, these comparisons with the neutral level do not guarantee the embodiment of Swarm Body; instead, they support that the level of embodiment is high in the \mbox{\textit{sparse}}.}

\subsubsection{Bone-Dynamic Could Realize Higher Level of Embodiment}
\label{subsubsec:vr_discussion_algorithm}
The sense of body ownership was found to be greatest for the \textit{bone-dynamic}, followed by the \textit{bone-static} and \textit{silhouette-dynamic}.
Similar to the sense of body ownership, the sense of agency showed a tendency that the \textit{bone-dynamic} results in the highest level followed by the \textit{bone-static}, and a significant difference between the \textit{bone-dynamic} and \textit{silhouette-dynamic} was found.
Thus, it is suggested that the \textit{bone-dynamic} algorithm results in the highest level of embodiment, while the \textit{silhouette-dynamic} algorithm results in the lowest.

The higher level of embodiment in \textit{bone}-based algorithms would come from the robot's ability to represent and respond to fingertip movements.
In the interviews, participants reported that they were more likely to perceive the swarm robot as a hand when the robot responded to their fingertip movements.
This suggests that the level of embodiment was improved when visual-motor synchronicity occurs even for local movements of the fingertips in addition to the whole hand movements.
As the \textit{bone}-based algorithm represents and responds to the participant's fingertip movements, it could enhance the level of embodiment through visual-motor synchronicity of fingertips.

In addition, similar to the \textit{dense} condition, the lower level of embodiment in the static condition may be due to the collisions, and oscillatory movements between robots are more likely to occur with the static assignment.
As such, the \textit{bone-dynamic} is considered to lead to the highest level of embodiment as it allows for visual-motor synchronization down to the fingertips while maintaining representation in various hand gestures and hand movements.

\section{Embodiment Experiment with Robots}
The VR experiment demonstrated that swarm robots \hl{could} be embodied, offering insights into their ideal behavior.
However, of course, there are differences between the VR environment and the real environment, and the use of real robots may have an effect on the results of the experiment.
Therefore, we designed a similar embodiment experiment with fewer factors while considering the VR experiment results to examine \hl{whether} similar embodiment characteristics \hl{can be} observed in real-world settings.

\subsection{Robot Implementation}
We developed custom-made swarm robots to conduct an embodiment experiment in the real world.
The robot design was inspired by Zooids ~\cite{legoc2016Zooids}, an open-source swarm robot platform, but the hardware and software were newly designed.

\subsubsection{Hardware Design}
\begin{figure}[h]
    \centering
    \includegraphics[width=\linewidth]{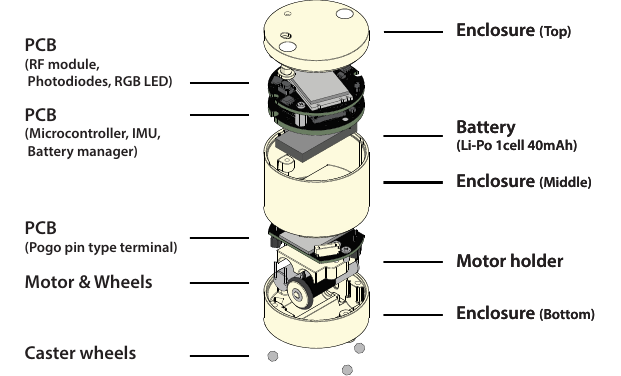}
    \caption{Exploded view of our custom-made swarm robot.}
    \Description{The figure shows the exploded view of our custom-made swarm robot. It is a cylindrical robot with three caster balls and two wheels at the bottom. From the top, the components are: a top enclosure with a button, two PCBs, a Li-Po battery, a middle enclosure, another PCB, a motor holder, a pair of motors and wheels, a bottom enclosure, and three caster balls.}
    \label{fig:robot_exploded}
\end{figure}
The VR experiment results suggest that the embodiment of swarm robots occurs for both 20~mm and 30~mm robots and that they show similar embodiment characteristics.
Also, owing to the limitations of currently available motor-based actuators and a communication module, the robot size cannot be reduced to 20~mm.
Therefore, we assumed that 30~mm robots could be used to conduct the embodiment experiment and designed our robot with that size.

The hardware design is illustrated in~\autoref{fig:robot_exploded}.
The robot parts include a microcontroller unit (STM32G071KBU6\footnote{\url{https://www.st.com/en/microcontrollers-microprocessors/stm32g071kb.html}} from STMicroelectronics), motor drivers (DRV8837DSGR\footnote{\url{https://www.ti.com/product/DRV8837/part-details/DRV8837DSGR}} from Texas Instruments), RF module (RF2401F20\footnote{\url{https://www.nicerf.com/item/nrf24l01-module-rf2401f20}} from NiceRF), motors with a 26:1 planetary gearbox (Pololu 2357\footnote{\url{https://www.pololu.com/product/2357}}), photodiode (PD15-22C/TR8\footnote{\url{https://everlighteurope.com/ir-detectors/2374/PD1522CTR8.html}} from Everlight Electronics), and a 40mAh Li-Po battery.

\subsubsection{Communication and Projector-based Tracking System}
\begin{figure}[h]
    \centering
    \includegraphics[width=\linewidth]{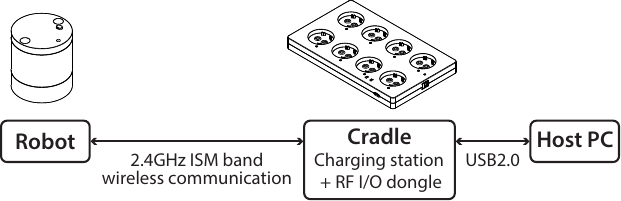}
    \caption{The robots communicate with the cradle through a 2.4 GHz ISM band wireless communication. The cradle communicates with the host PC with USB 2.0.}
    \Description{The figure shows the communication system. The robot is connected to a cradle, which works as a charging station and an RF I/O dongle. The cradle is connected to the host PC via USB 2.0.}
    \label{fig:robot_communication}
\end{figure}

The communication between the robots and the host computer is described in \autoref{fig:robot_communication}.
The methods are similar to the ones used for the Zooids.
The robots and cradles are each equipped with an RF module and communicate through a 2.4GHz ISM band wireless communication.
A projection-based localization system used for the Zooids was used to track the robots.
A high-speed projector (DLP LightCrafter 4500\footnote{\url{https://www.ti.com/tool/DLPLCR4500EVM}} from Texas Instruments) was used to project a sequence of gray-coded patterns onto the table; the two photodiodes on the robot received the projected coded-pattern light, and the microcontroller of the robot decoded its pattern into position information.
Then, the robot calculated its orientation from the positions of two photodiodes and broadcasted its position and orientation information to the host computer.

\subsubsection{Simulation-Based Robot Control}
The robots are controlled based on a simulation using the framework described in \autoref{sec:framework}.
In particular, we employed the same simulation for the real-world experiment as was used in the VR experiment.
Subgoal position generation and assignment are conducted \hl{based on tracked hand data}, and the robot positions are given by an RVO simulation with nonholonomic constraints.
Real robots are commanded to move to the current simulation robot positions every 100 ms.
In this manner, the robots can obtain incremental subgoal positions along their paths.
Then, the robot controls the rotation of the wheels according to the control law described in \autoref{sec:appendix} to reach the subgoal position.

\subsection{Participants}
A total of 10 participants (4 males and 6 females; 28.89 $\pm$ 13.83 (SD) years old) participated in the experiment.
Participants were recruited through a social media post.
All the participants were unaware of the purpose of the experiment, had normal or corrected vision, and were right-handed.
The participants signed a consent form regarding the experiment and were compensated with approximately \$16 on Amazon gift cards.
The ethics review board approved this study.

\subsection{Apparatus and Setup}
\begin{figure}[h]
    \centering
    \includegraphics[width=\linewidth]{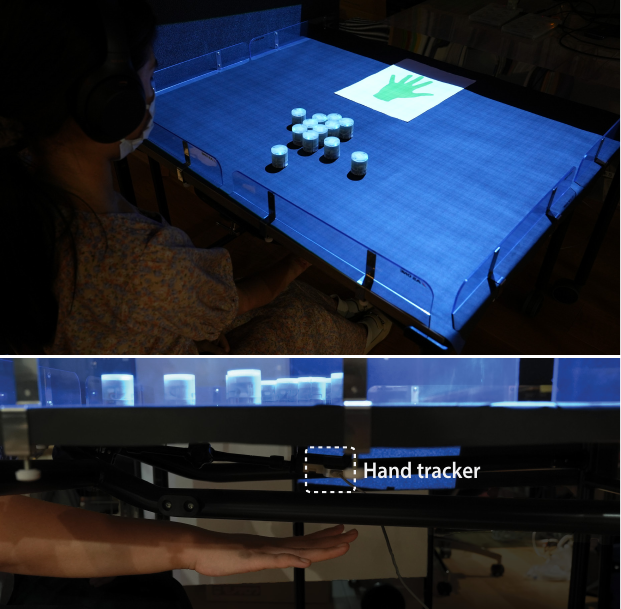}
    \caption{The experiment setup of the real-world robot study. The participant is wearing noise-canceling headphones and is controlling the swarm robots on the table with their right hand under the table (top). The hand is tracked with the hand tracker located under the table (bottom).}
    \Description{The experiment setup of the real-world robot study. The participant is wearing noise-canceling headphones and is controlling the swarm robots on the table with their right hand under the table (top). The hand is tracked with the hand tracker located under the table (bottom). The target hand sign printed on a piece of paper is placed on the table.}
    \label{fig:robot_experiment_setup}
\end{figure}

\autoref{fig:robot_experiment_setup} displays the experimental setup, which is similar to the VR experiment.
In this study, a hand tracker (Leap Motion Controller 2\footnote{\url{https://leap2.ultraleap.com/leap-motion-controller-2/}} from Ultraleap) was used to track hands instead of Meta Quest 2 to create the system without an HMD.
The hand tracker is located under the table, as shown in the \autoref{fig:robot_experiment_setup} (bottom).

\subsection{Experiment Design and Conditions}
This experiment was similar to the VR experiment.
\hl{However, the robot size was fixed at 30~mm, followed by a 3 $\times$ 3 factorial design.}
The independent variables examined were density (\textit{sparse}, \textit{medium}, and \textit{dense}) and subgoal position generation and assignment algorithm (\textit{bone-static}, \textit{bone-dynamic}, and \textit{\textit{silhouette-dynamic}}).
All variables were within the subject.

The task was similar to that in the VR experiment, which involved reaching a target with a specified hand shape.
One difference was that in the VR experiment, the targets were positioned in the right and left fronts, but in the real-world experiment, the target position was limited to the front, and accordingly, the number of tasks per trial was halved to four (\ie, four hand signs).
This is because the swarm robots' battery capacity would not hold a charge until the end of the experiment, with eight tasks per trial.

In addition, instead of the targets automatically (dis)appearing on the desk in the VR, the experimenter manually placed and removed the targets printed on paper on the desk.
To avoid the potential slipping of the robots on the target paper, the task was changed from moving the swarm robots to fit in the target to moving the swarm robots to a specified position in front of the target and making a specific hand shape.

\subsection{Measurements}
The same subjective measurements as those used in the VR experiment (\ie, the modified embodiment questionnaire and NASA TLX) were used to evaluate the sense of body ownership, sense of agency, and cognitive load.

\subsection{Procedure}
This procedure is similar to that used in the VR experiments.
\hl{Practice trials were conducted before the experiment to reduce the learning effects.}
After signing the consent form, participants started the practice trials (one under \textit{sparse} and \textit{bone-dynamic} and another under \textit{dense} and \textit{bone-dynamic}).

During the main experiment, the participants wore headphones with white noise and put their hands under the table to control the swarm robots.
The participants were instructed to move the swarm robots representing their right hand under the table to the starting area at the beginning of each task.
When all the robots returned to the starting area, the experimenter specified a hand sign by posting handshapes on paper on the desk.
The participants were instructed to move their swarm robot hand forward with a specified hand sign.
The hand sign sheet was removed after five seconds, and the participants moved their hands back to the starting area.
Participants repeated the reaching task and completed the questionnaire nine times.

To control the interference effect of arm fatigue, the participants were asked to ensure that they were not fatigued prior to each task.
A five-minute break was provided after the fifth task.
After the ninth questionnaire response, the participants answered a demographic questionnaire, and a semi-structured interview was conducted for approximately five to ten minutes.
The entire experiment took approximately one hour.

\subsection{Results}
\begin{figure}[h]
    \centering
    \includegraphics[width=\linewidth]{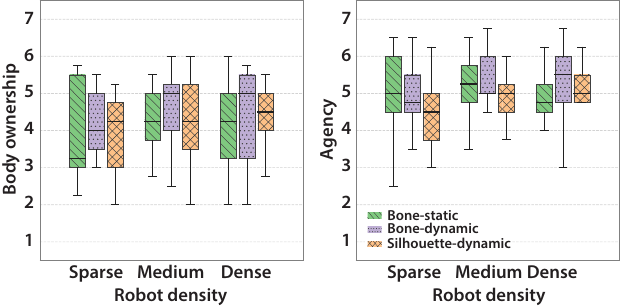}
    \caption{The body ownership and agency scores obtained in the robot experiment for each density and subgoal position generation and assignment algorithm.}
    \Description{The figure shows the box plot of the body ownership and agency scores obtained in the robot experiment for each density and subgoal position generation and assignment algorithm.}
    \label{fig:robot_bo_agency}
\end{figure}
\begin{figure}[h]
    \centering
    \includegraphics[width=\linewidth]{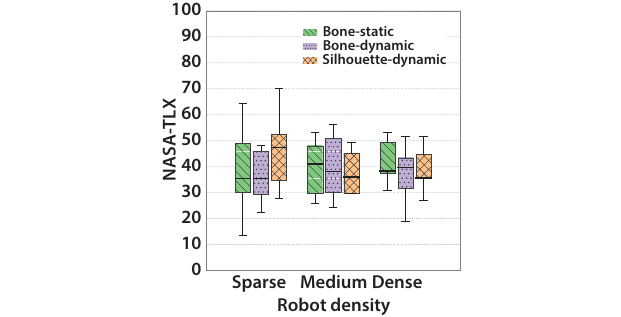}
    \caption{The NASA Task Load Index obtained in the robot experiment for each density and subgoal position generation and assignment algorithm.}
    \Description{The figure shows the NASA Task Load Index obtained in the robot experiment for each density and subgoal position generation and assignment algorithm.}
    \label{fig:robot_tlx}
\end{figure}

\hl{The body ownership score, agency score, and task load index were calculated for each combination of robot density and subgoal position generation and assignment algorithm.}
The hand tracker \hl{did not work properly and frequently lost track during the experiment of one of the participants, owing to the reflection of its own infrared light at} the bottom of the table.
As this might have strongly affected their data, we excluded them from the results and analysis.
The results are shown in \autoref{fig:robot_bo_agency} and \autoref{fig:robot_tlx}.
Similar to the analysis of the VR experiment results, we performed an ART procedure followed by a three-factor, two-way repeated-measures ANOVA with holm correction for the body ownership score, agency score, and task load index.

\subsubsection{Body Ownership}
ANOVA revealed a significant main effect of density ($F(2, 64)=3.31, $\hl{ $p=.043$}$, \eta_{p}^{2}=.09$).
No significant interaction was found between the density and algorithm factors.
Therefore, contrast tests were performed on the density and algorithm factors \hl{using the holm-corrected ART-C}.
The contrast test on the algorithm factor showed no significant differences in body ownership score between any of the conditions.
The test on the density factor showed that the \textit{medium} condition led to a significantly higher body ownership score than the \textit{sparse} condition~(\hl{$p =.047$, cohen's $d=0.676$}).

\subsubsection{Agency}
ANOVA revealed a significant main effect of density ($F(2, 64)=3.55, $\hl{ $p=.034$}$, \eta_{p}^{2}=.10$) and a moderate main effect of algorithm ($F(2, 64)=2.89, $\hl{ $p=.063$}$, \eta_{p}^{2}=.08$).
No significant interaction was found between the density and algorithm factors.
Therefore, contrast tests were performed on these factors using the holm-corrected ART-C.
The contrast test on the algorithm factor showed no significant differences in agency score between any of the conditions, but there was a trend that the \textit{bone-dynamic} condition led to a higher agency score than the \textit{silhouette-dynamic} condition (\hl{$p=.068$, cohen's $d=0.635$}).
The contrast test on the density factor also showed no significant differences in agency score between any of the conditions, but there were trends that \textit{medium} condition led to a higher agency score than the \textit{sparse} condition (\hl{$p=.052$, cohen's $d=0.665$}) and that the \textit{dense} condition led to a higher agency score than the \textit{sparse} condition (\hl{$p=.072$, cohen's $d=0.583$}).

\subsubsection{Cognitive Load}
ANOVA revealed no significant main effects of density and subgoal position generation and assignment algorithm.
No significant interactions were observed.
Therefore, contrast tests were performed on the density and algorithm factors using the holm-corrected ART-C.
The contrast tests on these factors revealed no significant difference in the task load index between \textit{sparse}, \textit{medium} and \textit{dense} conditions, as well as between the \textit{bone-static}, \textit{bone-dynamic}, and \textit{silhouette-dynamic} conditions.

\subsubsection{Semi-Structured Interview}
A few common items were identified during the interview at the end of the experiment.
\hl{Four} participants reported that the swarm robots felt like their hands in some trials, while the other five participants, including the participant under the poor hand tracking condition, reported that the robots were following their hands rather than being their own hands.
One of the five participants noted that it was difficult to distinguish between the hand signs, making it feel like a mass following the hand.
\hl{The other participant reported that the robots sometimes felt like their hand, but they were not fully convinced that it was their hand.}
Three participants mentioned that the control was intuitive when the robot was positioned for each fingertip.

Eight participants mentioned the influence of density on the embodiment.
Six stated that the dense robots felt more like \hl{the hand or something they could control better}; one preferred medium and sparse, and one preferred sparse.
One of those who preferred the dense condition stated that the dense swarm presented a sense of oneness and coherence.
Another participant reported that the swarm robots were embodied regardless of their density.

\subsection{Discussion}
To investigate the influence of swarm robot-unique factors on embodiment in real-world settings, we conducted a psychophysical experiment with the real swarm robots we developed, analyzing three aspects: sense of body ownership, sense of agency, and cognitive load.
The results were also compared with those of the VR embodiment experiment in \autoref{sec:vr_study}, discussing the unique characteristics of the real-world system's effect on swarm robot embodiment.

\subsubsection{Embodiment of Swarm Body with Bone-Dynamic Algorithm}
\label{subsubsec:embodimentBoneDynamic}
No significant difference was found between any of the algorithm conditions.
\hl{When compared with the neutral level (4 point rating in the 7-point Likert scale for body ownership and agency; 50 point rating in the 100-point NASA TLX for cognitive load),} the \textit{bone-dynamic} condition resulted in significantly higher body ownership and agency scores, and a significantly lower cognitive load score than the neutral levels with large effect sizes (body ownership: \hl{$p=.042$, cohen's $d=0.751$; agency: $p=.000$, cohen's $d=2.354$; cognitive load: $p=.000$, cohen's $d=1.722$, where the neutral levels were the null hypotheses}).
\hl{Thus, the questionnaire responses suggest that the participants felt a higher sense of embodiment for the \textit{bone-dynamic} algorithm than the neutral state in which participants neither deny or affirm whether they feel the robots as their body parts or as if they are in control of the robots.
This is consistent with the VR study finding that the \textit{bone-dynamic} condition achieved a higher level of embodiment than neutral levels.
This is also consistent with the interview responses from three participants that they could control swarm robots intuitively when the robots were positioned for each fingertip (\ie, bone-based algorithms).
The interview responses further suggest that some of the participants felt swarm robots as they would their hands in some trials though we cannot tell which algorithm conditions they talk about.
Overall, these questionnaire and interview results indicate that Swarm Body was possibly embodied in some trials, and if so, the embodiment probably occurred with the \textit{bone-dynamic} algorithm}.

\subsubsection{Shift in Preference toward Denser Swarm}
\label{subsubsec:shift}
\hl{When comparing the size conditions, t}he \textit{medium} density condition resulted in a significantly higher body ownership score than the \textit{sparse} condition.
The \textit{medium} and \textit{dense} conditions \hl{tended to result} in a higher agency score than the \textit{sparse} condition.
These results are consistent with the interview results that six out of eight participants who mentioned the influence of density preferred denser conditions.
However, these results seem to be inconsistent with the VR experiment's result that the \textit{sparse} condition led to a higher sense of body ownership than the \textit{dense} condition.
\hl{
There are three possible causes for this shift in preference for denser conditions in the real world: the increased importance of visual similarity, sense of accomplishment, and collisions.
}

First, visual similarity may be more important than understanding the correspondence between the body parts and robots in the embodiment in real-world settings.
As discussed in \autoref{subsubsec:vr_discussion_density} and \autoref{subsubsec:vr_discussion_algorithm}, sparse swarm robots with bone-based algorithms achieved a higher level of embodiment by offering a better understanding of the finger-robot correspondence through a simpler hand representation.
However, as suggested in~\cite{argelaguet2016role}, the visual similarity of an object to a hand affects its level of embodiment.
This effect might become more dominant in real-world settings.
This hypothesis was supported by the participants' comments that the dense swarm had a greater sense of oneness and coherence and that the denser the robot, the more it felt like a hand shape.

Second, the sense of agency may be influenced by the expected amount of effort required to move the object to be embodied.
Four participants reported a stronger feeling of controlling the robots and a stronger sense of accomplishment in the \textit{dense} condition.

Third, real-world systems have more collisions between robots, which might have reduced the level of embodiment in the \textit{sparse} condition.
As discussed in \autoref{subsubsec:vr_discussion_density}, the lack of collision\hl{s likely} contributed to the high level of embodiment in the \textit{sparse} condition in the VR study.
Real-world systems cause more collisions owing to the robot position errors between the simulation and the real world.
As a result, our real-world system had collisions even in the \textit{sparse} condition, which potentially decreased the level of embodiment.
\hl{It is also important to note that some robots turned off when they were stuck during the real-world study.
This is because our control program turns off a robot when the torque applied on its motors exceeds a certain threshold to protect the motors.
When that happened, the experimenter quickly turned them on, but this could affect the study results.
}

Thus, the participants' preferences shifted toward denser swarm robots, and the levels of embodiment in the \textit{dense} and \textit{medium} conditions were relatively higher in the real-world study.
\hl{
And the shift was possibly caused by the increased importance of visual similarity, a sense of accomplishment, and collisions.
This finding is somewhat limited by the fact that the preference shift might come from increased collisions which is partly dependent on our implementation.
However, as most real-world swarm robots have more collisions than their simulations, our finding is valuable for designing real-world embodied swarm robots though the degree of this shift may vary.
In addition, several consistent trends were observed throughout the VR and real-world experiments.
For example, higher embodiment levels were reported when the robots were located at the fingertips, and a lower cognitive load was measured in the \textit{sparse} condition}.

\section{Applications}
Swarm Body expands the design space of tangible and embodied interaction, \hl{offering unique characteristics such as robustness, flexibility, and scalability to the human body.}
The components can move without geometrical constraints other than collision with each other.
Additionally, although the current implementation requires a projector, the robot itself is not anchored to a specific environment; \hl{thus}, our system is versatile and can be used in a variety of locations, including ordinary desks.

\hl{The main application of Swarm Body is physical telepresence, where embodied swarm robots facilitate} physical interaction with remote people and environments, as shown in~\autoref{fig:teaser} (right).
The operator can control Swarm Body projected on their table as if manipulating their own hands.
The other person can physically interact with \hl{the operator through Swarm Body.}

Our physical telepresence system is inspired by Physical Telepresence Workspace by Leithinger~\etal~\cite{leithinger2014Physical}, \hl{particularly in} the physical representation of the user's hand using hand sensing and spatially-aligned visual feedback.
Our work extends their interaction capabilities \hl{through swarm robots characteristics, such as swarm splits, mergers, transfers, and obstacle avoidances.
Below, we outline scenarios that showcase new interaction opportunities in physical telepresence enabled by the characteristics of swarm robots.
}

\subsection{Multipliable Body}
\begin{figure}[h]
    \centering
    \includegraphics[width=\linewidth]{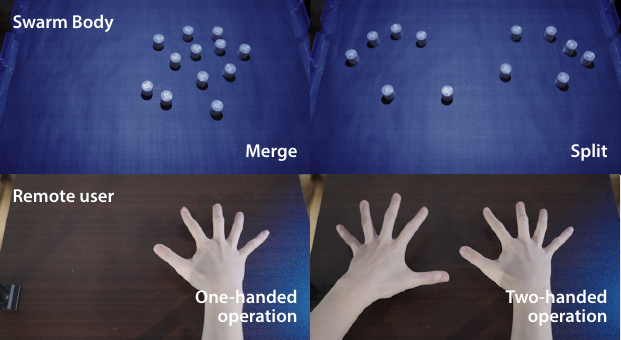}
    \caption{The remote user initially uses all the robots with their right hand (left). Then, they split them into two to use both of their hands (right).}
    \Description{The left figure shows the user's right hand and the corresponding swarm consisting of twelve robots. The right figure shows the user's two hands and the robots that correspondingly split into two groups. Each group consists of six robots.}
    \label{fig:app_divide_merge}
\end{figure}

\hl{
Swarm Body can split from a single swarm into multiple swarms, each representing different body parts, and then merge back into a single swarm.
These splits and mergers allow the user to adjust the number of independent swarms and the number of robots constituting each swarm as shown in~\mbox{\autoref{fig:app_divide_merge}}.
This enables the user to seamlessly switch between one- and two-handed telepresence in a single interface.
For example, when organizing a desktop remotely, the user can employ both hands to efficiently collect objects and then marge the robots to the dominant hand for precise organization.
The user can duplicate one hand into two, enabling the performance of two similar tasks simultaneously through parallel embodiment as in~\mbox{\cite{takada2022Parallel}}.
Additionally, unlike existing embodied robots that require one system for each user, Swarm Body supports multiple users manipulating the robots through a single interface.
For example, while one remote user engages in an activity such as rolling a ball with a local person, another remote user can allocate half of the robots to form a new hand and participate.
}

\subsection{Form-Giving to Transformable Body}
\begin{figure}[h]
    \centering
    \includegraphics[width=\linewidth]{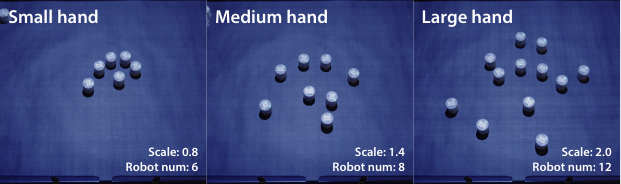}
    \caption{Swarm Body represents hands at various scales, allowing the user to experience different affordances of the environment.}
    \Description{The figures demonstrate how swarm robots can represent hands in different sizes. The left figure shows six robots representing a small hand. The middle figure shows eight robots representing a slightly larger hand. The right ifgure shows twelve robots representing an even larger hand.}
    \label{fig:app_transform}
\end{figure}

\hl{The malleability of Swarm Body enables the transformation of the swarm into various body parts of different sizes and unconventional forms (\eg, a small hand, elongated fingers, and a tentacle-like limb) (\autoref{fig:app_transform}).
This transformation provides enhanced interaction freedom while preserving embodiment features, such as intuitive swarm manipulation.
For example, one can experience the affordance of objects from a child's perspective by interacting with them using Swarm Body, which simulates a smaller hand, as in} \cite{nishida2020handmorph}\hl{.
Swarm Body can extend its fingers or transform them into tentacle shapes to reach and grasp objects at a distance or in narrow gaps.
Similar to how pixels on a screen represent a range of embodied avatars, Swarm Body physically embodies diverse avatars through its ability to transform.
}

\subsection{Adaptability to the Environment}
\begin{figure}[h]
    \centering
    \includegraphics[width=\linewidth]{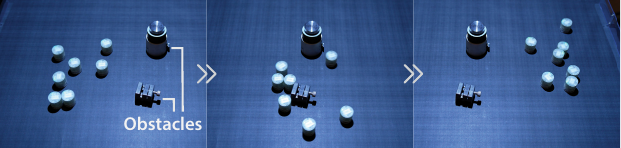}
    \caption{Swarm Body moves from left to right while passing through obstacles.}
    \Description{The figure demonstrates how swarm robots avoid obstacles on a table. The left figure shows seven robots in the scissors hand sign moving toward two obstacles. The middle figure shows the robots passing by the obstacles by slightly changing their formation. The right figure shows the robots having passed the obstacles and forming the scissors hand sign.}
    \label{fig:app_obstacle}
\end{figure}

\hl{Swarm Body adapts to its environment by avoiding obstacles, resizing, or transforming itself as needed.}
\hl{As shown in \mbox{\autoref{fig:app_obstacle}}, our control method introduced in~\mbox{\autoref{sec:framework}} enables the swarm robots to not only mimic the body movements but also avoid obstacles.}
In a tabletop environment with obstacles, the user can interact with the environment without the need to intentionally avoid the obstacles.
For example, when picking up a pen on the other side of a mug, the user can reach for it as if the mug did not exist.
\hl{We believe that Swarm Body could develop into a system that seamlessly adapts to various settings, easing interactions even in cluttered spaces giving it the potential to eliminate physical barriers in interactions, enabling smoother engagements than what our own bodies can typically achieve.
}

\subsection{Emotional Haptic Notification}
\begin{figure}[h]
    \centering
    \includegraphics[width=\linewidth]{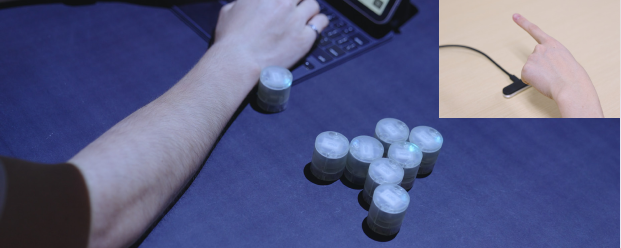}
    \caption{The Swarm Body user lightly taps the person with the finger to casually draw the person's attention.}
    \Description{The figure shows the user's hand and the corresponding swarm robots touching a person working on their computer. The user is tapping the person in the arm with the index finger.}
    \label{fig:app_haptic}
\end{figure}

\hl{
Swarm Body allows the user to exert horizontal forces on remote objects or individuals in an embodied manner.
This further expands the design space of physical telepresence with vertical actuation previously explored by Leithinger~\mbox{\etal~\cite{leithinger2014Physical}}.
Its embodiment aspect can also introduce intuitive and affective interactions, taking advantage of swarm characteristics in swarm user interface (SUI).
}
\hl{Swarm Body} achieves haptic communication by touching people~\autoref{fig:app_haptic}.
Specifically, they can naturally get a person's attention or express emotions to an intimate partner with various forces and touch.
For example, by gently tapping a person's arm engrossed in desk work, the robot can capture the person's attention and initiate communication.
Although previous studies have explored haptic \hl{feedback} using swarm robots~\cite{kim2019swarmhaptics}, our \hl{system} enables haptic feedback with the embodiment of the user.
\hl{
Thus, Swarm Body has the potential to haptically mediate the emotions and intentions of the user to the notified person. 
}

\subsection{Gesture Presentation}

\begin{figure}[h]
    \centering
    \includegraphics[width=\linewidth]{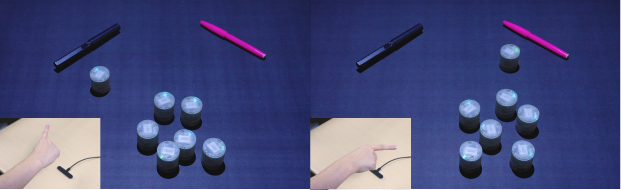}
    \caption{Swarm Body conveying the user's gestures. The user pointed at the black pen \hl{(left)}, then is now pointing at the pink pen \hl{(right)}.}
    \Description{The figures show the user's hand and the corresponding swarm robots on a table with a black pen and a pink pen. The user pointed at the black pen on the left, then is now pointing at the pink pen on the right.}
    \label{fig:app_gesture}
\end{figure}

Another interaction modality of Swarm Body is vision.
When embodied as hands, swarm robots can communicate through gestures with a physical presence in remote environments, as illustrated in~\autoref{fig:app_gesture}.
\hl{A remote individual can utilize physical gestures during their online presentations to boost engagement.
In a remote collaboration scenario, a remote user can point to specific objects or convey simple reactions.
For example, a remote craft instructor can point to the tools the students need to use at each step, direct their hand movements, and send a physical thumbs-up reaction upon task completion.
}
\section{Limitations and Future Work}
Our study did not comprehensively investigate the embodiment characteristics of swarm robots and their applications.

\subsection{More Extensive Investigation on Design Parameters}
Our investigation focused on a subset of the factor levels that should be examined to understand the embodiment of swarm robots on the hand.
Therefore, further studies should explore other possible factor levels.
For example, it is possible to investigate the embodiment characteristics of robots smaller than 20 mm in size or under even denser conditions than the $dense$ condition in the current work.
\hl{Studies in VR on these conditions may show how much embodiment is possible in theory.
The obtained level of embodiment for this theoretical condition will be a baseline when evaluating real-world systems.}

\hl{We did not examine some design parameters and complex conditions to make the experiments feasible.
Future research should explore various parameters including the robot's latency, speed, acceleration, color, and shape.
Additionally, dynamic changes of the parameters seem to be promising approaches.
For example, applying the bone-based algorithm to the fingertips and a silhouette-based algorithm to the other parts of the hand could integrate the advantages highlighted in our discussion.
Also, robots in a swarm can have different sizes, shapes, densities, and functions and change them dynamically as Li~\mbox{\etal~demonstrated~\cite{li2019Particle}}.
Therefore, further investigation on the embodiment characteristics and applications of such swarm robots and algorithms is expected.}

\subsection{Recognition as a Body Part by an External Observer}
\hl{
While we revealed the embodiment characteristics of swarm robots for the operator, we did not investigate whether an external observer could recognized the robots as someone's body parts.
In our preliminary testing, an observer was able to differentiate hand signs, although they were aware that the robots were representing a hand in advance.
Since visual feedback is the only information source for an observer, a silhouette-based subgoal generation algorithm, which reduced the embodiment level for the operator in our study, might enhance their recognition of the robots as a hand.
Further study on how an external observer recognizes the swarm robots is expected.
}

\subsection{Beyond Tabletop Robots and Embodiment of the Hand}
\hl{Swarm robots moving in 3D space, such as swarm drones, should be explored.
Although some of our methods and findings will be applicable to them, swarm robots moving in a 3D space have unique design parameters (\eg, subgoal positions on the skin surface only vs. those on the entire volume, including the interior of the hand).
This exploration also provides insight into how dimensional matching of the user's hand movements to the robot's subgoal formation affects a level of embodiment.
If 3D formations give a higher level of embodiment, then restricting the user's hand movements to the 2D plane (\ie, avoiding supinations and pronations) may also improve the level of embodiment in our 2D system.
In addition, their 3D formation and movements will open up a further interaction design space.
For example, embodied 3D swarm robots could pick up objects or enter space that is not accessible for wheeled swarm robots.
}
\hl{Note that the projection-based robot localization method used in our study could be adapted to estimate the position and orientation of such swarm robots in 3D space~\mbox{\cite{Hiraki2018Projection}}. 
These can be determined by solving the Perspective-n-Point problem, which involves considering the sensors mounted on the robots as points.}

The lack of investigation on body parts besides hands also limits this study.
We focused on hands as they are the most commonly used body parts for interaction with the environment.
Although we revealed the embodiment characteristics of swarm robots for the hand, it remains unclear if the same tendencies apply to other body parts.
We believe that our findings in the hand can be used to verify future studies on other body parts.
\hl{
Moreover, to expand the design space of Swarm Body, future research focusing on the effects of embodiment in body parts that differ from the user's actual body size and shape could be highly beneficial. 
This would open up new capabilities and applications for Swarm Body, allowing the user to manipulate objects at micro or macro scales with larger or smaller hands or using different shapes such as tentacle-like limbs.
}

\subsection{Exploration of the Applications}
Although we have presented several application scenarios, their effectiveness has yet to be evaluated.
In future work, we plan to showcase our applications through interactive demonstrations and videos, thereby collecting direct feedback from participants.
\hl{We also expect that this feedback will allow us to discover new application possibilities of Swarm Body beyond our initial scope.}


\subsection{Swarm Control Algorithm Dedicated for Embodied Behavior}
Lastly, collisions and misalignments between our robots might have affected the participants' evaluation of embodiment.
During the embodiment experiments in both VR and the real world, the experimenter sometimes observed collisions and misalignments of the robots.
These undesirable behaviors potentially come from conditions, such as a large robot size or a high density, that make it difficult for swarm robots to avoid collisions while following a hand.
\hl{As mentioned in~\mbox{\autoref{subsubsec:shift}}, our control program shuts down the robot in the case of excessive torque to protect the motors.
This might affect the participants' sense of embodiment, leading to a different result from the VR study, even though the experimenter turned them on immediately.
In other words, the participants' preference toward the denser swarm robots might be the characteristics of our real-world implementation not our approach.
However, our real-world experiment and the interview responses provided some consistent and generalizable insights, such as a higher level embodiment coming from fingertip representations and a lower cognitive load in sparse conditions.
}
To address the collision issues, a specialized swarm control algorithm tailored for embodied behavior could potentially reduce collisions and enhance the embodied experience.
\hl{Further psychophysical embodiment experiments with a control algorithm with fewer collisions will help us understand whether the differences in our VR and real-world study results are due to the approach or our implementation.
It will also deepen our understanding on the embodiment characteristics we observed in both VR and real-world settings (\ie, a higher level of embodiment coming from fingertip representations and a lower cognitive load in sparse conditions).}

\section{Conclusion}
We proposed a new embodied system concept, \emph{embodied swarm robots}, a group of robots collectively acting as a human body part.
We presented a framework for the embodiment of swarm robots, investigating their characteristics in both VR and real-world environments.
Our results offer two key insights into the embodiment of swarm robots.

\begin{enumerate}
    \item \hl{Swarm robots are likely to be embodied in VR and real-world scenarios using a suitable algorithm and density though there are some individual differences}.
    \item The choice of swarm body control algorithm influences the level of embodiment, impacting both the visual-motor synchronicity of fingers and the frequency of robot collisions.
\end{enumerate}

Additionally, we explored applications of our system, demonstrating how embodied swarm robots can enrich tangible and embodied interactions between humans and the environment.



\begin{acks}
We thank all of our participants for their time and invaluable feedback.
Our special thanks go to Zendai Kashino, Masahiko Inami, and researchers at The University of Tokyo's Information Somatics Lab for their discussions.
We also thank Karakuri Products, Inc. and Noura Howell and her students at Georgia Institute of Technology for their wonderful help.
This work is partially supported by JST AIP Acceleration Research JPMJCR23U2, Japan.
\end{acks}

\bibliographystyle{ACM-Reference-Format}
\bibliography{sample-base}

\appendix

\section{Appendix}
\subsection{Swarm Robot Control for Following Subgoal Position}
\label{sec:appendix}
The robot receives the subgoal (target) position from the host computer via RF communication and moves to follow it.
It is desirable for the robot to follow a smooth path to the subgoal position.
In controlling wheeled robots, using a Bézier, spline, or cross-oid curve as the path is common.
This curve passes through the current and subgoal positions to avoid sudden changes in angular velocity.
However, due to the limited computational resources of the robot's microcontroller, it has been difficult to implement control that sequentially calculates and follows these curves.
On the other hand, since the robot can get its absolute position information with the projection-based method, and the subgoal position is updated by the host computer every 100 ms, the distance between the current position and the subgoal position is considered to be close, and we thought that sudden changes in angular velocity would be rare even on a path connecting these two points by a straight line.
Thus, we implemented a control model that follows the straight path.

\begin{figure}[h]
    \centering
    \includegraphics[width=\linewidth]{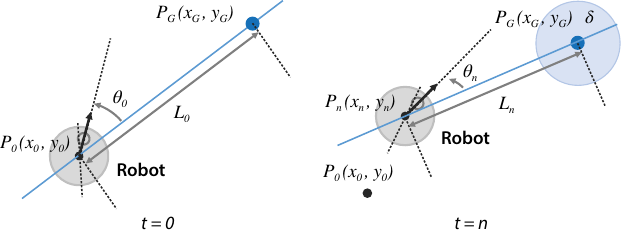}
    \caption{Concept of robot control model for following goal position.}
    \label{fig:robot_control_model}
\end{figure}
The concept of the control model is shown in \autoref{fig:robot_control_model}.
Let $P_0 (x_0, y_0)$ be the robot's initial position at time $t = 0$ given the subgoal (target) position and $P_G (x_G, y_G)$ the subgoal position.
Next, define $\overrightarrow{P_0 P_G}$ as the vector connecting points $P_0$ and $P_G$, with $L$ ($L = L_0$ at $t = 0$) as the distance between them, and $\theta$ ($\theta=\theta_0$ at $t = 0$) as the angle between the robot's direction of motion and the vector $\overrightarrow{P_0 P_G}$. (Counterclockwise is defined as positive.)
In this case, the following law controls the velocities $V_l$ and $V_r$ of the robot's left and right motors.
\begin{equation*}
\begin{split}
    V_l &= V - \Delta V\\
    V_r &= V + \Delta V\\
    V &= K_L L\\
    \Delta V &= K_\theta \theta + K_{\dot{\theta}} \dot{\theta},
\end{split}
\end{equation*}
where $K_L, K_\theta, K_{\dot{\theta}}$ are constants that represent the control gain.
Based on the above control law, $\overrightarrow{P_0 P_G}$, the distance $L_n$, and the angle $\theta_n$ are updated from the robot's current position $P_n (x_n, y_n)$ to the subgoal position, and $V_l$ and $V_r$ are calculated and output.
However, if this control law is followed, the point $P_G$ is theoretically unreachable and will never fully converge, since $L = 0$ and the velocity will converge to 0 as the robot approaches $P_G$.
Therefore, the robot is judged to have converged when it is within a certain distance $\sigma$ from $P_G$, and the robot stops at that point.
In addition, the minimum velocities $V_{l,min}$ and $V_{r,min}$ are set for $V_l$ and $V_r$, respectively.
The maximum speed of motors $V_{max}$ is determined by the maximum speed value of the slower of the two motors.
This is based on the actual measured speeds of the left and right motors as an upper limit.
This can be written in the formula as follows:
\begin{equation*}
    \begin{gathered}
        V_l = 
        \begin{cases}
            V_{l,min} & \text{if $V_l < V_{l,min}$}\\
            V_{max} & \text{if $V_l > V_{max}$}\\
            V_l & \text{otherwise.}
        \end{cases}
        , V_r = 
        \begin{cases}
            V_{r,min} & \text{if $V_r < V_{r,min}$}\\
            V_{max} & \text{if $V_r > V_{max}$}\\
            V_r & \text{otherwise.}
        \end{cases}
        , \\V_{max} = 
        \begin{cases}
            V_{l,max} & \text{if $V_{r,max} \geq V_{l,max}$}\\
            V_{r,max} & \text{if $V_{r,max} < V_{l,max}$}
        \end{cases}
        ,
    \end{gathered}
\end{equation*}
where $V_{l,max}$ and $V_{r,max}$ are the maximum speed value of left and right motors.

If a new subgoal position $P_G$ is given before convergence, $P_G$ is updated, and the subgoal following continues.
Here, the dimension (unit) of the calculated $V_l$ and $V_r$ is [mm/s], but the dimension of the PWM duty of the voltage value, which is a control value that can be input to the motor, is [rpm].
Therefore, the calculated $V_l$ and $V_r$ are converted to the PWM duty value by the following formula:
\begin{equation*}
\begin{split}
    Duty_l &= f( V_l )\\
    Duty_r &= g( V_r ),
\end{split}
\end{equation*}
where $Duty_l$ and $Duty_r$ are the PWM values of left and right motors, and $f(x)$ and $g(x)$ are functions corresponding to the PWM value calculated by the robot's calibration mode.
We confirmed that this control rule can be used to control the robot to successfully reach the goal position even with a microcontroller with limited computational resources.







\end{document}